\documentclass[12pt,epsfig]{article} 
\input{epsf.tex}
\usepackage{amssymb,epsfig}
\newcommand{\One}{1\kern-4.5pt1}
\newcommand{\lapprox}{\raisebox{-0.5ex}{$\ 
\stackrel{\textstyle<}{\textstyle\sim}\ $}}
\newcommand{\gapprox}{\raisebox{-0.5ex}{$\ 
\stackrel{\textstyle>}{\textstyle\sim}\ $}}
\begin{document}

\addtolength{\baselineskip}{0.20\baselineskip}

\rightline{SWAT/04/419}

\hfill December 2004

\vspace{48pt}

\centerline{\Huge Lattice Study of Anisotropic QED$_{3}$ }

\vspace{18pt}

\centerline{\bf Simon Hands  and
\bf Iorwerth Owain Thomas }

\vspace{15pt}

\centerline{\sl Department of Physics, University of Wales Swansea,}
\centerline{\sl Singleton Park, Swansea SA2 8PP, U.K.}
\vspace{24pt}


\centerline{{\bf Abstract}}

\noindent
{\narrower
We present results from a Monte Carlo simulation of non-compact lattice 
QED in 3 dimensions
on a $16^3$ lattice in which an explicit anisotropy between $x$ and $y$ hopping
terms has been introduced into the action. This formulation is inspired by
recent formulations of anisotropic QED$_3$ as an effective theory of the
non-superconducting portion of the cuprate phase diagram, with relativistic 
fermion degrees of freedom defined near the nodes of the gap function on the
Fermi surface, the anisotropy encapsulating the different Fermi
and Gap velocities at the node, 
and the massless photon degrees of freedom reproducing the dynamics
of the phase disorder of the superconducting order parameter.
Using a parameter set corresponding in the isotropic limit to 
broken chiral symmetry (in field theory language) or a spin density wave
(in condensed matter physics language),
our results show that the renormalised anisotropy, defined in terms of the ratio
of correlation lengths of gauge invariant bound states in the $x$ and $y$ 
directions, exceeds the explicit anisotropy $\kappa$ 
introduced in the lattice action,
implying in contrast to recent analytic results 
that anisotropy is a relevant deformation of QED$_3$.
There also appears to be a chiral symmetry restoring phase transition at 
$\kappa_c\simeq4.5$, implying that the pseudogap phase persists down to
$T=0$ in the cuprate phase diagram.
}




\vfill
\newpage
\section{Introduction}

\begin{figure}[htb]
\bigskip\bigskip
\begin{center}
\epsfig{file=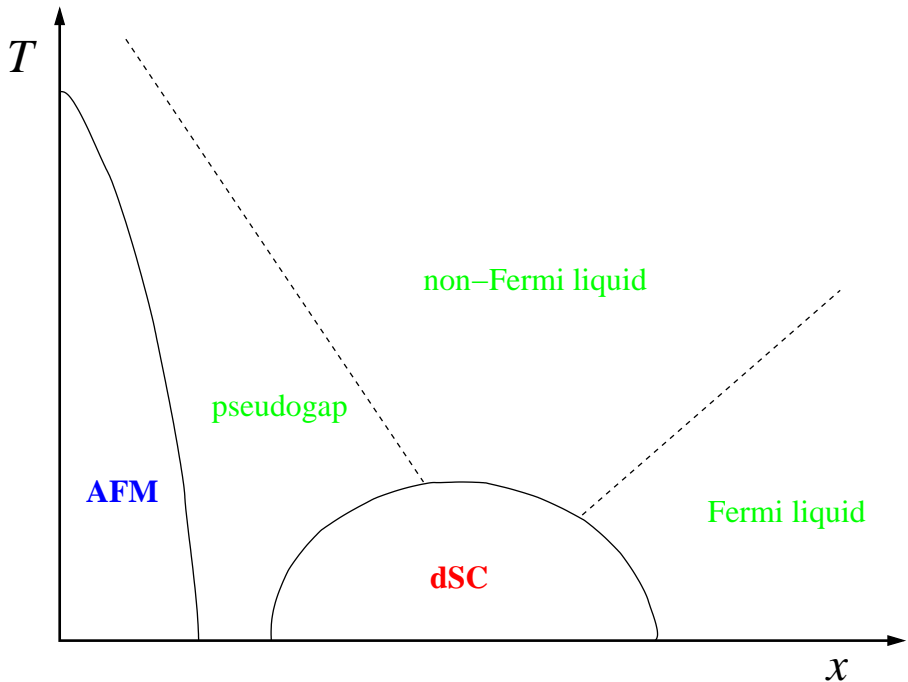, width=12cm}
\end{center}
\caption{(Color online) Schematic phase diagram for cuprate superconductors}
\label{fig:phase}
\end{figure}

The phase diagram of the superconducting cuprate compounds in the $(x,T)$ plane,
where $x$ denotes the doping, or fraction of holes per CuO$_2$ unit, continues
to be the object of much study, both experimental and theoretical.
A schematic
version is shown in Figure~\ref{fig:phase} \cite{Varma}.
Around $x\sim0.2$, so-called {\em optimal doping}, there is a superconducting
dSC phase 
extending to temperatures as high as $T\sim50$K. The superconductivity is
believed to be an essentially two-dimensional phenomenon, being confined 
to the CuO$_2$ planes, and the gap function $\Delta(\vec k)$ is characterised by
$d$-wave symmetry, thus having two pairs of nodes on the one-dimensional Fermi
surface. For $x\approx0$ the compound is an insulating anti-ferromagnet, and
as $x$ increases in this AFM phase  the order smoothly
evolves into a spin-density
wave characterised by a wavevector $\vec K$ whose magnitude decreases with $x$.

In some sense the ``normal'' phase is more strange. While in the over-doped 
regime the behaviour is that of a normal metal, namely a conventional Fermi
liquid, as $x$ is decreased  unusual non-Fermi liquid 
behaviour manifests itself via non-standard $T$-scaling 
of transport coefficients
such as resistivity and themal conductivity. More mysterious still is the 
``pseudogap'' behaviour observed in the under-doped region; 
as one moves out of the 
dSC phase in the direction of increasing $T$ or decreasing $x$,  
studies of the 
spectral density distribution function at fixed momentum
show the 
quasiparticle pole of the superconductor 
(correponding to a well-defined excitation of energy $\Delta$
above the Fermi energy) 
diminish in strength, but the magnitude of the energy gap $\vert\Delta\vert$
remain non-zero even in the non-superconducting region. 
This spectral depletion can persist up to $T\sim150$K \cite{Timusk}.   

It should be stressed that while AFM and dSC both have well-defined order
parameters, and hence are separated from the rest of the phase diagram by solid
lines in Figure~\ref{fig:phase}, the status of the dashed lines separating the 
``normal'' phase into three regions is currently much less clear. Nonetheless,
there have been several attempts to formulate a theoretical description 
of the pseudogap region. A particularly interesting programme,
starting from established properties of the dSC phase,
derives an effective theory which resembles QED in 2+1 dimensions,
but having spatial anisotropy in the covariant derivatives
\cite{Franz:2002qy,Herbut:2002yq}. The starting point is the Bogoliubov -- de
Gennes model for $d$-wave quasiparticles in the dSC phase, which in Euclidean
metric (corresponding to the imaginary time formalism in many body theory)
has action
\begin{eqnarray}
S&=&T\sum_{\vec k,\sigma,\omega_n}\biggl[(i\omega_n-\xi_{\vec k})
c^\dagger_\sigma(\vec k,\omega_n)c_\sigma(\vec k,\omega_n)\nonumber\\
&-&{\sigma\over2}\left(\Delta(\vec k)c^\dagger_\sigma(\vec
k,\omega_n)c^\dagger_{-\sigma}(-\vec k,-\omega_n)
-\Delta^\dagger(\vec k)c_\sigma(\vec
k,\omega_n)c_{-\sigma}(-\vec k,-\omega_n)\right)\biggr],
\label{eq:BdG}
\end{eqnarray}
where $c^\dagger,c$ are creation and annihilation operators for electrons with 
spin $\sigma=\pm1$, and $\omega_n=(2n-1)\pi T$ are the allowed Matsubara
frequencies. The function $\xi_{\vec k}$ is the energy of a free quasiparticle,
which thus vanishes for $\vec k$ on the Fermi surface, and $\Delta(\vec k)$ is 
the gap function, which can be thought of as a self-consistent pairing field.
The requirement of $d$-wave  symmetry implies that 
$\Delta(\vec k)=0$ at four special
node momenta $\vec k=\pm\vec K_1,\pm\vec K_2$ with $\vec K_1.\vec K_2=0$. 
If we choose axes 
such that $\vec K_1\parallel \hat x$, $\vec K_2\parallel\hat y$, 
and write $\vec k=\vec K_i+\vec q$,
then in the vicinity of the ``1'' nodes it is possible to linearise as
\begin{equation}
\xi_{\vec k}=v_Fq_x+O(q^2)\;\;\;\;;\;\;\;\;\Delta(\vec k)=v_{\Delta}
q_y+O(q^2)
\end{equation}
and near the ``2'' nodes as
\begin{equation}
\xi_{\vec k}=v_Fq_y+O(q^2)\;\;\;\;;\;\;\;\;\Delta(\vec k)=v_{\Delta}
q_x+O(q^2),
\end{equation}
where the parameters $v_F$ and $v_\Delta$ are the {\em Fermi} and {\em Gap}
velocities respectively.

The next stage is to define a 4-spinor $\Psi_i$ at the node $i$:
\begin{equation}
\Psi_i^{\mbox{tr}}(\vec q,\omega)=\left(c_+(\vec k,\omega), c^\dagger_-(-\vec
k,-\omega), c_+(\vec k-2\vec K_i,\omega), c^\dagger_-(-\vec k+2\vec
K_i,-\omega)\right).
\end{equation}
The association of different spinor components with different points in
$k$-space is well-known to workers in lattice QCD familiar with the staggered
fermion formulation \cite{Golterman:1984cy}.
It is now possible to recast the low energy limit of
the kinetic term of
(\ref{eq:BdG}) in relativistic garb:
\begin{eqnarray}
S=\int d^2r\int_0^\beta dt
\bar\Psi_1[\gamma_0\partial_t&+&v_F\gamma_1\partial_x
+v_\Delta\gamma_2\partial_y]\Psi_1+\nonumber\\
\bar\Psi_2[\gamma_0\partial_t&+&v_F\gamma_1\partial_y
+v_\Delta\gamma_2\partial_x]\Psi_2+O(\partial^2,\Psi^4)
\label{eq:rel}
\end{eqnarray}
where $\beta\equiv T^{-1}$, 
$\bar\Psi=\Psi^\dagger\gamma_0$, and the $4\times4$ traceless hermitian 
Dirac $\gamma$-matrices
obey $\{\gamma_\mu,\gamma_\nu\}=2\delta_{\mu\nu}$. It is important to stress
that there is no reason {\em a priori} for the 
anisotropy encapsulated in the ratio $v_F/v_\Delta\equiv\kappa$ to be
negligble in real cuprates; a value as high as $\sim10$ 
is reported in \cite{Chiao}.
In addition, $\kappa$ increases with doping fraction $x$ \cite{Sutherland}.

To give the nodal fermions interactions, 
it is necessary to discuss the reason for 
the loss of superconducting order. The hypothesis is that the gap function can 
be written as $\Delta=\Delta_0(\vec k)e^{i\theta(\vec r)}$, where $\Delta_0$ is
real, and that
superconductivity is destroyed because the phase field $\theta$ becomes
disordered. In two spatial dimensions non-trivial phase disorder arises 
through the accumulation of {\em vortices}, that is, point dislocations around
which $\oint\nabla\theta.d\vec\ell=2n\pi$ \cite{BKT}. 
The dSC $\to$ pseudogap transition 
is thus supposedly driven by vortex condensation, which preserves
$\vert\Delta\vert\sim\Delta_0$ but ensures $\langle e^{i\theta}\rangle=0$.
Now, it is possible to exploit the gauge symmetry of (\ref{eq:BdG}) to
absorb phase fluctuations of $\Delta$ into the phases of $c,c^\dagger$ and hence
$\Psi$. 
We would thus seek a theory of phase fluctuations for the $\Psi$ fields.
However, since $\Delta$ represents a doubly-charged Cooper pair field,
it is impossible to do thus while maintaining single valuedness, since
the phase of  $\Psi$ would change by only $\pi$ on circling a vortex.
The solution proposed by Franz and Te\v sanovi\'c \cite{FTgauge}, is to
partition the vortices of any particular configuration $\{\theta\}$
into two groups A \& B, 
and then
to associate the phase $\theta_A(\vec r)$ 
associated with A vortices exclusively with the spin-up
electrons, and that of the B to spin-down. It can then be shown 
\cite{Franz:2002qy,Herbut:2002yq} that the
relativistic $\Psi$ fields of (\ref{eq:rel}) couple minimally to the
vector-valued
difference field $a_\mu={1\over2}\partial_\mu(\theta_A-\theta_B)$, which thus
acts as an effective ``photon'' in the gauge-invariant action which results
from replacing $\partial_\mu$ in (\ref{eq:rel}) by the covariant derivative
$D_\mu=\partial_\mu+ia_\mu$.

It immediately follows from gauge invariance, which implies that the vacuum
polarisation tensor correcting $\langle a_\mu(p)a_\nu(-p)\rangle$
has the transverse form $(p^2\delta_{\mu\nu}-p_\mu p_\nu)\Pi(p)$,
that the $a_\mu$ excitations do not receive a mass due to 
quantum corrections from the $\Psi$ fields.
Further 
arguments have been advanced \cite{Franz:2002qy,Herbut:2002yq} to suggest that 
fluctuations in the $a_\mu$ field are themselves 
governed by the action of 2+1$d$
electrodynamics
\begin{equation}
S_{phot}={1\over 2g^2}\int d^2r\int_0^\beta dt
(\partial_\mu a_\nu-\partial_\nu a_\mu)^2,
\label{eq:sphot}
\end{equation}
with the coupling $g$ related to the 
diamagnetic susceptibility $\chi$ via $g\sim\chi^{-{1\over2}}$, or in field
theoretic terms to the dual order parameter $\tilde\Phi$ for
vortex condensation via $g\sim\langle\tilde\Phi\rangle$. 

One way of
understanding this is that in the absence of magnetic monopoles (which
in 2+1$d$ are instantons) which can act as
a source or sink of flux, vorticity is a topologically-conserved charge.
When the ground state is such that $\langle\tilde\Phi\rangle\not=0$, the 
U(1) global symmetry for which vorticity is a Noether charge, i.e. the timelike
component of a conserved current $\tilde V_\mu$,  is spontaneously
broken, resulting in a massless boson in the spectrum via Goldstone's theorem.
However, in 2+1$d$ this Goldstone boson is kinematically equivalent to 
the photon, as can be seen via, eg, the PCAC-like relation \cite{Kovner:1990nr}
\begin{equation}
\langle 0\vert\tilde V_\mu\vert \mbox{1 photon}, \vec p\rangle\propto p_\mu.
\end{equation}

As a result of these considerations, 
it is natural to unite (\ref{eq:rel}) and (\ref{eq:sphot})
and postulate a relativistic field theory,
 QED$_3$ with $N_f=2$ flavours of 
nodal fermion $\Psi$, 
as the appropriate effective
action for low energy long wavelength excitations in the pseudogap phase --
the photons $a_\mu$ interacting with the $\Psi$ with
effective electric charge $g$. 

QED$_3$ is an asymptotically-free theory, which means that it becomes more
strongly interacting as energy scales decrease, ie. as length scales grow. 
The infra-red behaviour of QED$_3$ has
long been a challenge to theory (see
\cite{HKS} for a brief review); in brief, the
issue is whether the chiral symmetry $\Psi\mapsto e^{i\alpha\gamma_5}\Psi$, 
$\bar\Psi\mapsto\bar\Psi e^{i\alpha\gamma_5}$ of (\ref{eq:rel}) (with
$\gamma_5\equiv\gamma_0\gamma_1\gamma_2\gamma_3$) is spontaneously broken 
by a parity-invariant  fermion  -- anti-fermion
condensate $\langle\bar\Psi\Psi\rangle\not=0$. This is believed to happen
if the number of flavours $N_f$ is smaller than some critical value $N_{fc}$,
which has been variously estimated  as taking values in the range 
$\lapprox{3\over2}$
to $\sim5$. The consequence is a dynamically generated fermion mass $\Sigma$; 
the determination of the exact value of $N_{fc}$, and the dynamically generated
ratios $\Sigma/g^2$ and $\langle\bar\Psi\Psi\rangle/g^4$ remain outstanding
problems in non-perturbative field theory.

To understand this issue in the context of cuprates it is necessary to return to
the original electron variables $c,c^\dagger$. If $N_{fc}>2$ then chiral
symmetry is broken at $T=0$ in the long-wavelength or continuum limit. This
translates into a non-vanishing value for 
$\langle\sum_{\sigma,i}\sigma\cos(2\vec K_i.\vec r)c^\dagger_\sigma(\vec r)
c_\sigma(\vec r)\rangle$, which is nothing but the order parameter for spin
density waves characterised by wavevectors $2\vec K_i$ \cite{Herbut:2002yq}. 
This implies that for 
sufficiently small $T$ there is a direct passage from AFM to dSC
without going through an intermediate strip of the
pseudogap phase, and a corresponding triple point at the intersection of AFM,
dSC and normal phases for some $T>0$. 
If, on the other hand, $N_{fc}<2$, then we expect the pseudogap phase
to persist all the way down to $T=0$, as sketched in Fig.~\ref{fig:phase}.
In either case, it may be possible to explain the non-Fermi liquid properties
in terms of a non-perturbatively 
large anomalous scaling dimension for $\Psi$ in the 
chirally-symmetric phase of QED$_3$ \cite{Franz:2002qy}.

The above discussion ignores the effects of the anisotropy
$v_F/v_\Delta\equiv\kappa>1$. This has been justified by analytic
treatments performed for small departures from isotropy
in the limit of large-$N_f$ \cite{Lee:2002qz,Vafek:2002jf},
where it is shown  anisotropy is irrelevant in the renormalisation group (RG)
sense. Specifically, 
Ref.~\cite{Lee:2002qz} uses a
Schwinger-Dyson approach in the large-$N_f$ limit to
study the behaviour of $\kappa$ under RG flow, finding
\begin{equation}
{d{\kappa_{ren}}\over ds}=-{32\over{5\pi^2N_f}}(\kappa_{ren}-1)
\end{equation}
where 
$s$ is the logarithm of the ratio of UV cutoff to physical momentum scales.
This implies 
that for weak anisotropy ($\kappa\gapprox1$) $\kappa_{ren}$ is 
driven to 1 under RG flow and hence $\kappa_{ren}-1$ is an 
irrelevant parameter, ie. $\frac{\kappa_{ren}-1}{\kappa-1}<1$.  
Hence, it is argued, predictions from isotropic QED$_3$ can be applied
directly to cuprates.

The purpose of the current paper is to examine this claim for arbitrary 
$\kappa$ and the ``physical'' case $N_f=2$. The theoretical tool we use is
lattice simulation of so-called non-compact QED, modelling the fermi degrees of
freedom as staggered lattice fermions. The lattice method is fully
non-perturbative, with systematically 
improvable errors due to a finite UV cutoff (the inverse
lattice spacing $a^{-1}$), and a non-zero IR cutoff 
(the inverse lattice size $L^{-1}$) of a
completely different nature to other approaches. The non-compact nature of the
model, to be defined more fully in Sec.~\ref{sec:model} below, has the effect
of suppressing  
monopoles (which appear as point singularities in the phase field $a_\mu$),
thus maintaining the masslessness of the photon
\cite{Polyakov:1976fu,Motrunich:2003fz}. Lattice simulations of isotropic
QED$_3$ have in the past been applied to the issue of $N_{fc}$: such attempts
have been hampered by large finite volume effects in the continuum limit
due to the massless photon, 
and to date the results are only able to suggest $N_{fc}>1$
\cite{HKS,Hands:2004bh}. In this study we work away from the continuum limit 
at a coupling $g$ sufficiently strong that we are confident chiral symmetry is
broken at $\kappa=1$, and then systematically increase $\kappa$.
Further details of the lattice model and our simulation are given in
Sec.~\ref{sec:model}. In Sec.~\ref{sec:numbers} we present numerical results,
and in particular present evidence firstly for a restoration of chiral symmetry
at some critical $\kappa_c$, and secondly for the renormalised anisotropy
$\kappa_{ren}(\kappa)>\kappa$, where $\kappa_{ren}$
is defined in terms of certain correlation lengths in differing directions, 
implying in contrast to the analytic results 
that anisotropy is a relevant
deformation of QED$_3$. Implications for cuprate superconductivity are 
discussed in Sec.~\ref{sec:discuss}.

\section{The Model and Simulation}
\label{sec:model}

The lattice formulation of QED with non-compact gauge fields and staggered
lattice fermions is described in detail in ref.~\cite{Hands:2004bh}.
The following is an $N$ flavour staggered fermion action for QED$_{3}$ with 
explicit $x-y$ anisotropy:
\begin{equation}
S =\sum_{i=1}^{N} \sum_{x,x'} a^3\bar{\chi_{i}}(x) M_{x,x'} \chi_{i}(x') 
+\frac{\beta}{2} \!\!\sum_{x,\mu<\nu}\!\! a^3\Theta_{\mu\nu}^{2}(x) 
\label{eqn:lattact}
\end{equation}

The fermion matrix $M_{x,x'}$ is defined as follows:
\begin{equation}
	M_{x,x'} =  {1\over2a}\sum_{\mu=1}^{3} \xi_{\mu}(x) 
[\delta_{x',x+\hat{\mu}} U_{x\mu} -
\delta_{x',x-\hat\mu}U_{x'\mu}^\dagger]
+ m\delta_{\mu\nu}  \label{eqn:fermion_matrix}
\end{equation}
with $\xi_{\mu}$ given by 
\begin{equation}
\xi_{\mu}(x) = \lambda_{\mu} \eta_{\mu}(x)
\end{equation}
and $\eta_{\mu}(x)=(-1)^{x_{1} + ... + x_{\mu-1}}$, where $x_1=x$, $x_2=y$ and
$x_3=t$,  
is the Kawomoto-Smit phase of the staggered fermion field. 
The $\lambda_{\mu}$ are 
anisotropy factors, to which we assign the following values: 
$\lambda_{x}=\kappa^{- \frac{1}{2}}$, $\lambda_{y}=\kappa^{\frac{1}{2}}$, 
$\lambda_{t}=1$.  The purpose of the phase factors is to ensure that in the 
isotropic limit $\lambda=1$ the action describes relativistic covariant
fermions \cite{Golterman:1984cy}.

If the photon-like degree of freedom $\theta_\mu(x)$ is defined on the link
connecting site $x$ to site $x+\hat\mu$, then
$U_{x\mu}\equiv \exp(ia\theta_{x\mu})$ in (\ref{eqn:fermion_matrix}) is 
the parallel transporter defining the gauge interaction with the fermions, 
and we have a non-compact gauge action given by 
\begin{equation}
	\Theta_{\mu\nu}(x) = {1\over a}[\Delta_{\mu}^{+}\theta_{\nu}(x)-  
\Delta_{\nu}^{+}\theta_{\mu}(x)]. 
\end{equation}
The dimensionless parameter $\beta$ is given in terms of the QED coupling
constant (ie. the ``electron charge'')
via $\beta\equiv1/g^2a$, where $a$ is the lattice spacing. It is
convenient to work wherever possible in units such that $a=1$.
As discussed above, 
we use a non-compact formulation of the gauge fields because flux symmetry is not preserved in compact U(1) formulations due to instanton formation, 
which causes the photon in such formulations to be 
massive \cite{Polyakov:1976fu,Motrunich:2003fz}.

If we restrict our attention to that portion of the action involving the fermion
fields, then we see that the introduction of the $\lambda_\mu$ factors has the 
effect, at least at tree level, of rescaling the lattice spacing in the various
directions as $a_x=\surd\kappa a$, $a_y=a/\surd\kappa$, $a_t=a$. In orthodox 
lattice QCD similar anisotropies 
are often introduced for technical convenience; 
eg. spectroscopy of highly excited states such as glueballs is considerably
more efficient if $a_t\ll a_x,a_y,a_z$ \cite{Morningstar:1997ff}. 
In this case to ensure Lorentz covariance of the continuum limit
it is important to check that all
terms in the lattice action are formulated with the same anisotropy, which
results in a fine-tuning problem once quantum corrections are
introduced. For instance, implementing this programme for the 
action (\ref{eqn:lattact}) would
require the introduction of separate gauge coupling constants $\beta_{xt}$,
$\beta_{yt}$, and $\beta_{xy}$, with a non-trivial constraint resulting from the
physical requirement that eg. the speed of light for photons is the same as that
for fermions. In the case at hand, though, the plaquette coupling $\beta$ is
defined the same in all three planes. 
It is important to stress that in this case
the $x-y$ anisotropy is {\em physical}, and that eg. the resulting ratio
$a_x/a_y$ is an observable to be determined empirically.
At tree level $a_x/a_y=\kappa$; in what follows (see
Sec.~\ref{sec:kren}) we define this ratio as the {\em renormalised} anisotropy 
$\kappa_{ren}$ and estimate it from the spatial 
decay of a mesonic correlation function. Rather than keep track of the various
lattice spacings, we prefer to think of
$\kappa$ as a parameter of the model which can be renormalised through quantum
corrections. This approach was pioneered in 
Ref.~\cite{Lee:2002qz}, where it was shown using large-$N_f$ arguments that 
$\kappa_{ren}<\kappa$ (note that in \cite{Lee:2002qz} 
the equivalent parameter is called $\lambda$,
and that our model sets their parameter $\delta$ to 1).
 
We set $N=1$, which yields $N_{f}=2$ fermion species in the continuum limit.
An algebraic transformation exists relating the 
single component staggered fields $\chi,\bar\chi$ to four-component continuum
spinors $\Psi$
\cite{Burden:1986by}, and in particular the chiral condensates
are related via 
$\langle\bar\chi\chi\rangle=\sum_i\langle\bar\Psi_i\Psi_i\rangle$. 
However, we note that in the simulations presented in this paper
we are working at a strong coupling ($\beta=0.2$), 
far from the continuum limit; 
this was done so that we could be reasonably confident that chiral symmetry is 
broken \cite{Hands:2004bh}, 
allowing us to examine the effects of anisotropy on the model's phase
stucture starting from the putative AFM phase.
A certain amount of caution is 
mandated in applying our results to the condensed matter-inspired QED$_{3}$ 
model (\ref{eq:rel},\ref{eq:sphot}), which is 
derived and justified in continuum terms.  Further caution is warranted as the 
flavour structure of (\ref{eqn:fermion_matrix}) does not entirely capture the theory
of \cite{Franz:2002qy,Herbut:2002yq};  
in the condensed matter-inspired theory (\ref{eq:rel}) the second 
flavour has a 
$v_{F}\gamma_1$ term in the $y$ direction and a 
$v_{\Delta}\gamma_2$ term in the $x$ direction so the two flavours have
opposite anisotropies, reflecting the fact that there
is no physical anisotropy in the original crystal: in our model by contrast, 
following the transformation to $\Psi,\bar\Psi$ variables the 
the 
velocity-$\gamma$ matrix structure of the first flavour would be repeated in the
second, so that there is an overall anisotropy.  
We expect however that enough
similarities between (\ref{eqn:lattact}) and the cuprate-inspired model
persist
for us  to make reasonable conjectures as to the behaviour of the 
latter system. The point will be discussed further below.

In order to aquire our results, 
we utilised a Hybrid Monte Carlo simulation of the 
action (\ref{eqn:lattact}) with even-odd partitioning on a $16^{3}$ 
lattice at $\beta=0.2$, 
simulating for anisotropies $1 \leq \kappa \leq 10$  at bare masses of $0.01 
\leq m \leq 0.05$. Further details can be found in \cite{Hands:2004bh}; here it
suffices to note that the algorithm generates representative  configurations of 
$\{\theta\}$ weighted according to the action (\ref{eqn:lattact}) in an 
{\em exact}, that is to say unbiased manner. It is in principle possible to
perform simulations with a lattice action corresponding more closely to the
anisotropy structure of (\ref{eq:rel}), but in this case simulations
would have to be performed with a Hybrid Molecular Dynamics algorithm, and
results would thus contain a systematic dependence on the timestep size used
\cite{Hands:2004bh}. This algorithm would then approximate 
the ``correct'' model via a
functional measure $[\mbox{det}M(\kappa)M(\kappa^{-1})]^{1\over2}$; 
however, away 
from the continuum limit it remains an unresolved issue whether the resulting
dynamics is that of a local Lagrangian field theory.

Around 1000 trajectories of mean length 1.0 
were generated for each data point, 
and acceptance rates were generally in the region of 70-80\% for $0.02 \leq m 
\leq 0.05$ and 60-70\% for $m = 0.01$ (where simulations were less efficient)
apart from $\kappa=10.00$, whose acceptance was over 80\%.

\section{Numerical Results}
\label{sec:numbers}

We performed measurements of the following parameters in our simulation: 
the mean 
gauge action $\langle{\beta\over2}\Theta_{\mu\nu}^2\rangle$ separately in each 
$\mu\nu$ plane, the chiral condensate $\langle\bar\chi\chi\rangle$, 
the longitudinal susceptibility 
$\chi_l\equiv\partial\langle\bar\chi\chi\rangle/\partial m$, 
and the pion correlator $C_\pi$ in each direction, from which we extracted the 
pion mass ($t$ direction) and the effective masses (also known as {\em inverse
screening lengths}) in both space 
directions $x$ and $y$.

\subsection{Plaquettes}
\label{sec:plaquettes}

\begin{figure}[tb]
\begin{center}
\begin{minipage}[c][7.8cm][c]{7.4cm}
\begin{center}
\epsfig{file=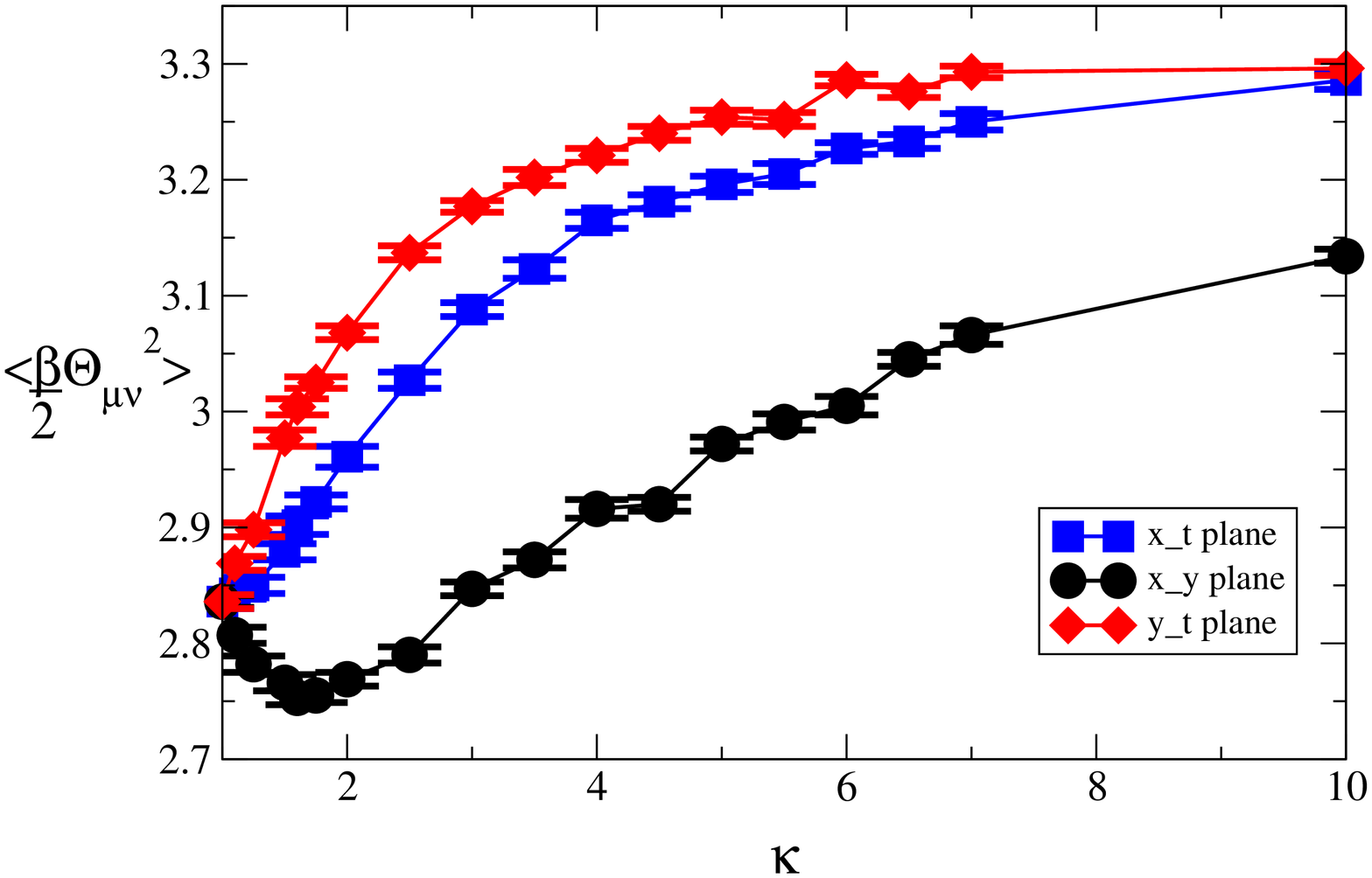, angle=270, width=7.8cm}\\[-1mm]
(a)
\end{center}
\end{minipage}
\begin{minipage}[c][7.8cm][c]{7.4cm}
\begin{center}
\epsfig{file=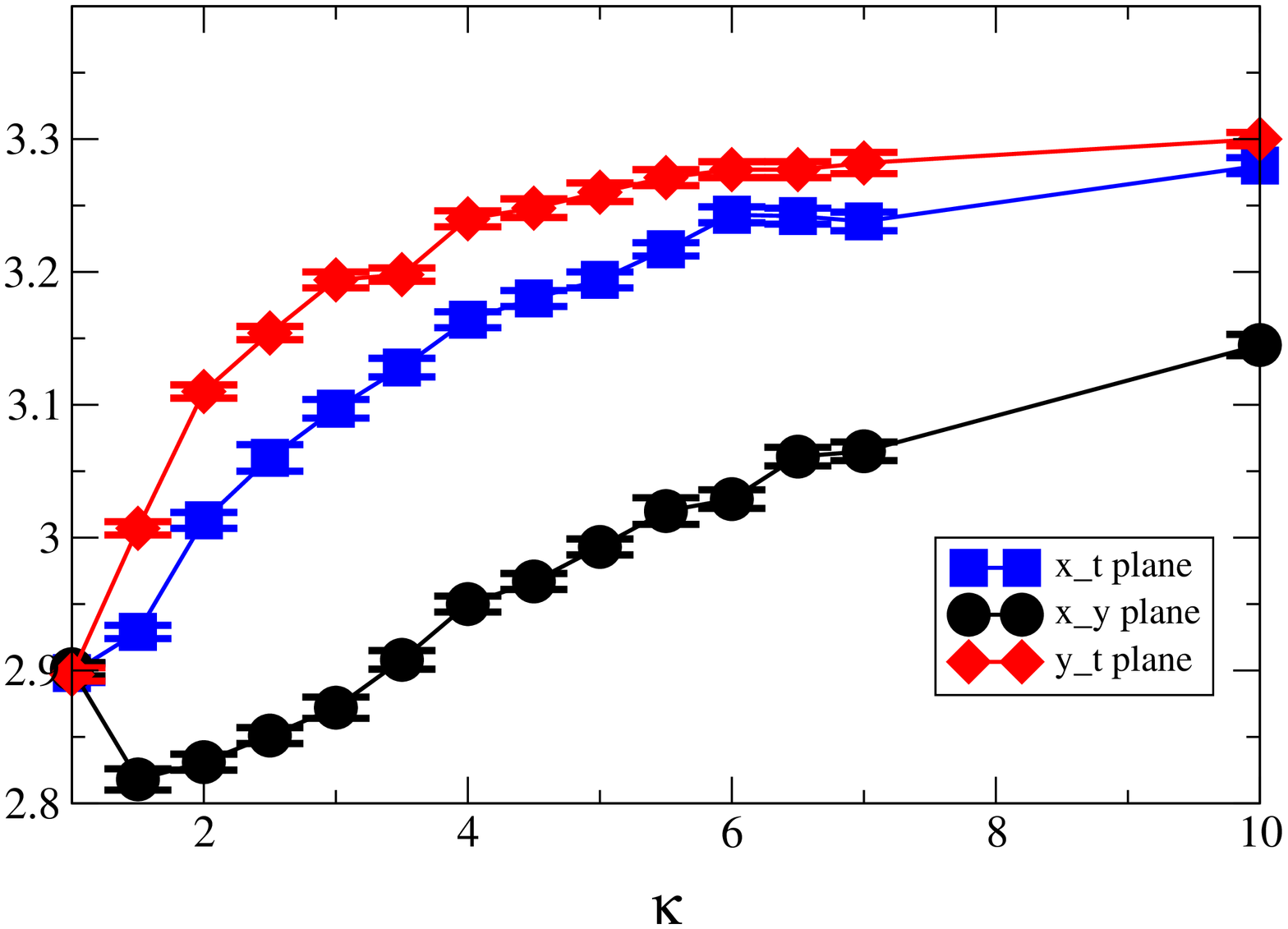, angle=270, width=7.8cm}\\[-1mm]
(b)
\end{center}
\end{minipage}
\caption{(Color online) Plaquette action for (a) $m=0.01$, and (b) $m=0.05$ as functions of
anisotropy $\kappa$}
\label{fig:plq}
\end{center}
\end{figure}
Fig.~\ref{fig:plq} shows average gauge action 
values for $\kappa$ between 1 and 10 for bare fermion mass
$m=0.01$ (a) and 0.05 (b).
This is an important observable in exposing dynamical
fermion effects: since the plaquette term $\propto\Theta^2$
in (\ref{eqn:lattact}) is
$\kappa$-independent, any anisotropy effect must be due to 
the effects of quantum corrections due to the fermion sector.

The general upward trend of the $x-t$ and $y-t$ plaquette values 
with $\kappa$ could plausibly be 
explained as a result of reduced 
screening of bare charge due to
quantum corrections. In perturbation theory the dominant screening process 
is known as
vacuum polarisation --- 
virtual light $\Psi\bar\Psi$ pairs decrease the effective value of $g$
and hence increase the effective $\beta$;
this implies that 
fluctuations $\langle\Theta^2\rangle$ 
are more strongly suppressed by the dynamics of
(\ref{eqn:lattact}). This can be seen by comparing the $m=0.05$ and $m=0.01$
data at $\kappa=1$. The increase of $\langle\Theta^2\rangle$ with $\kappa$
would therefore suggest that light $\Psi\bar\Psi$ 
pairs become {\em less} important as anisotropy
increases. It will prove difficult, at first sight, 
to reconcile this observation
with results of Sec.~\ref{sec:chiral}.

That the value of the average $x-t$ plaquette is consistently less than that in the 
$y-t$ plane can be explained as being due to the overall $x-y$ anisotropy
of (\ref{eqn:lattact}); 
in the condensed
matter model (\ref{eq:rel}) anisotropic effects 
should cancel between the two flavours. Accordingly, we can interpret the 
$\lapprox4$\%
mismatch between $\langle\Theta_{xt}^2\rangle$ and $\langle\Theta_{yt}^2\rangle$
in Fig.~\ref{fig:plq} as some measure of the systematic error in our treatment.

However, the $x-y$ plane plaquettes are markedly different. 
First, the behaviour 
is non-monotonic -- we observe two regimes, one for 
$1<\kappa \lapprox 1.5$, and another for  $\kappa \gapprox 1.5$.  
Within the latter, we have behaviour consistent with the other two planes, 
but for small $\kappa$ the mean plaquette value 
{\em decreases} as $\kappa$ is increased.  
Secondly, the relative splitting between $x-y$ and $x-t$, $y-t$ is much larger, 
$O(10\%)$. In this case there is no symmetry argument to suggest that this
splitting should vanish for dynamics based on the condensed matter model
(\ref{eq:rel},\ref{eq:sphot}) and indeed, the
approximately quadratic behaviour for small $\kappa$ suggests that some residual
anisotropy effect should survive even in an $x-y$ symmetrised model, in which
effects linear in $\kappa$ might be expected to cancel between the two flavours
\cite{ref:Herbfoot}.

\subsection{Restoration of Chiral Symmetry}
\label{sec:chiral}
\begin{figure}[htb]
\bigskip\bigskip
\begin{center}
\epsfig{file=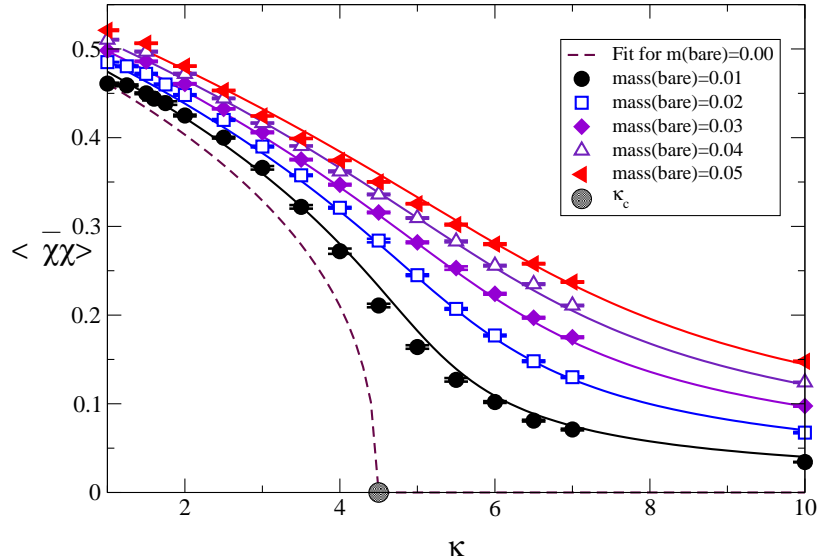, angle=270, width=12cm}
\end{center}
\caption{(Color online) Chiral condensate $\langle\bar\chi\chi\rangle$ versus $\kappa$. Lines
denote fits to the equation of state (\ref{eq:eos})}
\label{fig:chiralgraph}
\end{figure}
The chiral condensate is defined in terms of the inverse fermion matrix:
\begin{equation}
\langle\bar{\chi}\chi\rangle=-{1\over V}{{\partial\ln Z}\over\partial m}=
{1\over V}\langle\mbox{tr}M^{-1}\rangle,
\end{equation}
and the longitudinal susceptibility as:
\begin{equation}
\chi_{l}=\frac{\partial\langle\bar{\chi}\chi\rangle}{\partial m}
={1\over
V}[\langle\mbox{tr}M^{-1}\mbox{tr}M^{-1}\rangle-
\langle\mbox{tr}M^{-1}\rangle^2-\langle\mbox{tr}M^{-1}M^{-1}\rangle].
\label{eq:lsusc}
\end{equation}
Anisotropy effects
observed within the fermion
sector should also be present in the symmetrised model (\ref{eq:rel}), 
although of course with
the roles of $x$ and $y$ reversed for the second flavour.

\begin{figure}[htb]
\bigskip\bigskip
\begin{center}
\epsfig{file=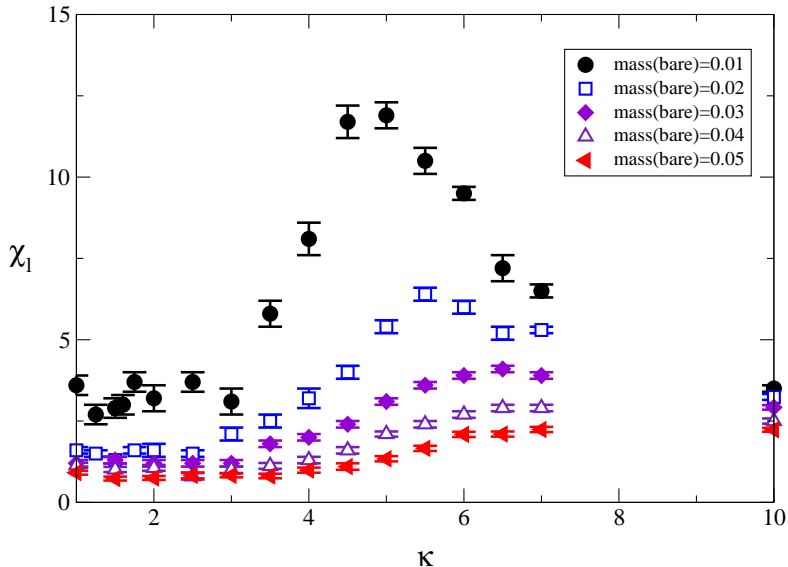, angle=270, width=12cm}
\end{center}
\caption{(Color online) $\chi_{l}$ versus $\kappa$.} 
\label{fig:longsusc}
\end{figure}

Figures \ref{fig:chiralgraph} and \ref{fig:longsusc} depict
$\langle\bar\chi\chi\rangle$ and 
$\chi_l$ for bare masses between 0.01 and 0.05 for various $\kappa$.  
Note first of all that for $\kappa\gapprox1$ $\langle\bar\chi\chi\rangle$ 
varies 
very little as $m$ decreases, and certainly extrapolates to a non-zero value 
as $m\to0$, implying the spontaneous breaking of chiral symmetry in this
limit. This is consistent with the behaviour observed at strong coupling in 
\cite{Hands:2004bh,HKS}.
Figs.~\ref{fig:chiralgraph} and \ref{fig:longsusc} both suggest
a chiral symmetry restoring phase transition 
as  
$\kappa$ is increased: a drop in the value of the chiral condensate in 
Figure \ref{fig:chiralgraph} (most pronounced for $m=0.01$) and, in 
Figure \ref{fig:longsusc}, a peak in the susceptibility which grows more 
prominent as the mass is decreased. 

We performed a fit of the $\langle\bar\chi\chi\rangle$ data
to a hypothetical equation of state of the 
form
\begin{equation}
m=A(\kappa-\kappa_c)\langle\bar\chi\chi\rangle^\rho
+B\langle\bar\chi\chi\rangle^\delta,
\label{eq:eos}
\end{equation}
which assumes a second order phase transition at $m=0$, $\kappa=\kappa_c$ with 
conventionally-defined critical exponents $\delta$ and
$\beta_{mag}=(\delta-\rho)^{-1}$ \cite{Hands:2004bh}.
We chose to fit 
values of $\kappa$ from 1 to 7, as there are too few points above $\kappa=7.00$
to give a good definition of the curve.  Fixing the values of the exponents 
$\rho$ and $\delta$ to 1 and 3 respectively gives us a mean field approximation;
we find $A=0.0710(2)$, $B=1.382(6)$ and $\kappa_{c}=5.018(7)$ with
a $\chi^{2}/d.o.f.$ of 72.  If $\rho$ and $\delta$ are allowed to vary, 
we obtain $A=0.103(1)$, $B=2.75(6)$, $\kappa_{c}=4.35(2)$, $\rho=1.297(9)$ and 
$\delta=3.99(3)$ with a 
$\chi^{2}/d.o.f.$ value of 51. 
The latter fit is plotted as solid lines 
in 
Fig.~\ref{fig:chiralgraph}, with the dashed lines denoting the equation of state
in the chiral limit.  
Despite the large values of $\chi^2$, 
these curves seem to describe the data reasonably well;  we 
conjecture that if a phase transition occurs, 
it will do so at $\kappa_{c}\approx 4.5$. 
A finite volume scaling study is needed before the order of the
phase transition, and hence the validity of (\ref{eq:eos}) can be established 
unambiguously.


While Figs.~\ref{fig:chiralgraph} and \ref{fig:longsusc} show clear
evidence of a phase transition across which $\langle\chi\chi\rangle$ decreases
dramatically, 
it is too early to draw conclusions regarding its precise nature.
For a second order transition, $\chi_l$
should diverge at the critical anisotropy in the chiral limit $m\to0$.
However, without a comparison of data from different volumes  a
first order transition cannot be excluded. 
In the current context the implications are profound:
a second 
order transition would lead us to expect $\langle\bar{\chi}\chi\rangle=0$ for 
$m=0$, $\kappa > \kappa_{c}$, implying $N_{fc}<2$ in this regime, whereas
if the transition is first order it remains
conceivable that $\langle\bar{\chi}\chi\rangle \neq 0$ is small but non-zero,
and hence $N_{fc}>2$.
Caution is required because 
simulations of isotropic QED$_{3}$ with $N_f=2$
cannot exclude a very small dimensionless condensate
$\beta^2\langle\bar\chi\chi\rangle<10^{-4}$ in the continuum limit \cite{HKS}, 
invisible on the scale of Fig.~\ref{fig:chiralgraph}, but nonetheless 
perfectly consistent with recent analytical estimates \cite{AW}.

\subsection{Pion Correlation Functions and Spectroscopy}
\label{sec:spectro}
In this subsection we focus on the correlation functions 
\begin{equation}
C_{\pi\mu}(x_\mu)=\sum_{\nu\not=\mu}\sum_{x_\nu}
\langle\bar\chi\varepsilon\chi(0)\bar\chi\varepsilon\chi(x)\rangle,
\label{eq:picorr}
\end{equation}
where the phase $\varepsilon(x)\equiv(-1)^{\sum_\mu x_\mu}$. 
In the isotropic limit $\kappa=1$ on a symmetric $L^3$ lattice
all the $C_{\pi\mu}$ coincide. When chiral symmetry is spontaneously 
broken it can be shown
that the correlator is dominated by one of $N^2$
pseudoscalar approximate Goldstone boson poles whose 
mass $m_\pi^2\propto m$. By analogy with particle physics we refer to such
states
as {\em pions}; in continuum notation they are interpolated by the operator
$\bar\Psi\gamma_5\Psi$ \cite{chiralfoot}. Now, in Euclidean
quantum field theory for sufficently large separation $\vert x_\mu\vert$ 
the correlator can generally be fitted by the form 
\begin{equation}
C_{\pi\mu}(x_\mu)=A(e^{- m_{\pi\mu} x_\mu} + e^{-m_{\pi\mu} (L_{\mu} -x_\mu)}),
\label{eqn:cosh}
\end{equation}
where $L_\mu$ is the extent of the lattice in the $\mu$ direction.
For $\mu=t$ the decay parameter $m_{\pi t}$ is the pion mass, ie. the excitation
energy to create a pion at rest, whereas for $\mu=x,y$
the corresponding quantities are identified as inverse screening lengths. Of
course, on an isotropic symmetric 
lattice corresponding to $T\approx 0$ all three coincide,
but in our system with explicit $x-y$ anisotropy, $m_{\pi x}$, $m_{\pi y}$ and
$m_{\pi t}$ are all distinct.

\subsubsection{Pion correlators in the time direction}
\label{sec:pionprop}

\begin{figure}[htb]
\bigskip\bigskip
\begin{center}
\epsfig{file=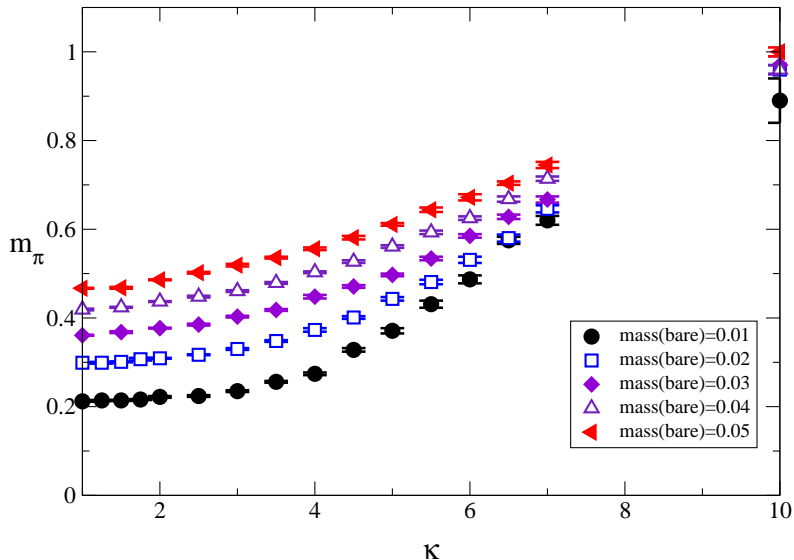, angle=270, width=12cm}
\end{center}
\caption{(Color online) $m_{\pi t}$ as $\kappa$ increases.} 
\label{fig:pimass}
\end{figure}
Figure \ref{fig:pimass} shows the variation of $m_{\pi t}$ as $\kappa$ is 
increased.  These values were extracted from the timeslice pion propagator 
(\ref{eq:picorr}) via 
least squares fitting to (\ref{eqn:cosh}).
Some caveats must be offered regarding this data: it is apparent that for 
the very lightest pions
the masses were too small for the lattice size (ie. $L\gg m_\pi^{-1}$ is not
satisfied), meaning that we could not use 
an effective mass plot in order to estimate the ideal fitting window for each
mass.
Instead, 
we chose the fit window that provided the best $\chi^{2}$ value and that 
produced a curve 
that passed through the error bars of as many of the propagator 
data points as possible.  Because of this, a table listing all the data points, 
their $\frac{\chi^{2}}{d.o.f}$ values and their fit windows is given below 
(table \ref{tab:tpimass}).  
Despite the uncertainty this procedure yields in the 
absolute values of $m_{\pi t}$, the {\em trends} found in the data
are
not artefacts of the fit window chosen, coinciding for masses where the 
fit window is more or less stable ($m=0.01$) as well as those where it is less 
so (eg. $m=0.05$).

It is clear from Figure \ref{fig:pimass} that $m_{\pi t}$ for all $m$ increases 
with $\kappa$ is increased, most dramatically for  $\kappa\gapprox4$, 
where the $m=0.01$ data show a perceptible kink.
It appears therefore that in the chiral limit $m\to0$ we have two 
regimes, one where $m_{\pi t}$ is relatively insensitive to anisotropy, and one 
where $m_{\pi t}$ increases approximately linearly with $\kappa$.  
It is tempting to identify the boundary
between these two
regimes with $\kappa_c$ of the last section
-- indeed, non-analytic behaviour across a phase transition should be
expected since the pion is a Goldstone mode 
of the broken phase $\kappa<\kappa_c$.  
The linear behaviour of $m_{\pi t}(\kappa)$ for all masses in the region 
$5 \lapprox \kappa \lapprox 7$ is not as yet understood.
  
The behaviour of pions at low energies can be described by the non-linear 
sigma model:
\begin{equation}
S_{NLSM}=\frac{f_\pi^2}{2}  
\int d^{3}x (\partial_{\mu} U)^{\dagger}(\partial^{\mu} U),
\end{equation}
where $U(x)\equiv\exp(i\pi(x)/f_\pi)$ is a unitary matrix of the chiral group
$G$ 
and the coupling $f_{\pi}$, known as the the pion decay constant, parametrises
the strength of pion self-interactions, which become weak in the limit $k\to0$. 

\begin{figure}[htb]
\bigskip\bigskip
\begin{center}
\epsfig{file=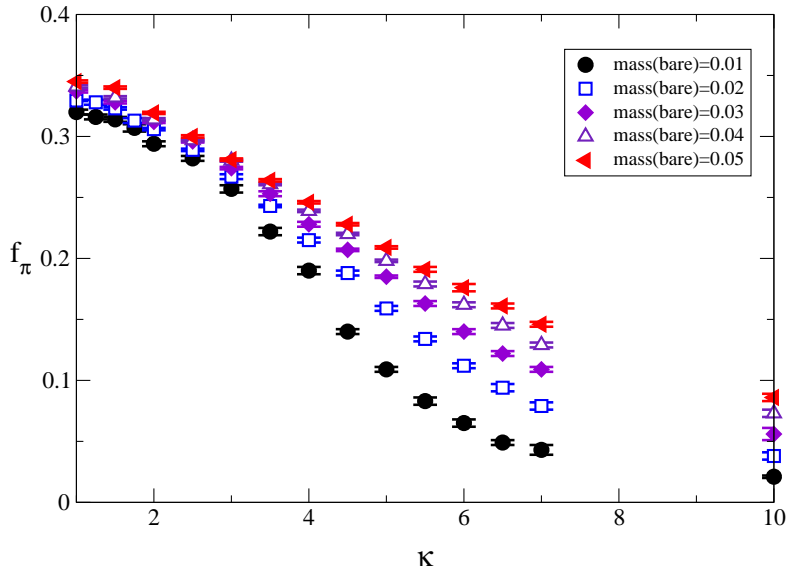, angle=270, width=12cm}
\end{center}
\caption{(Color online) $f_{\pi}$ as $\kappa$ increases.} 
\label{fig:fpi}
\end{figure}
We may calculate $f_{\pi}$ for various 
$\kappa$ by relating $m_{\pi}$ and $\langle\bar{\chi}\chi\rangle$ using the 
Gell-Mann-Oakes-Renner relation \cite{Gell-Mann:1968rz},
\begin{equation}
m_{\pi}^{2} f_{\pi}^{2}= m\langle\bar{\chi}\chi\rangle
\label{eqn:GMORrln}
\end{equation}
The results are plotted in Figure \ref{fig:fpi}.  
They track the $\langle\bar{\chi}\chi\rangle$ 
results of Figure \ref{fig:chiralgraph} very closely; this is perhaps 
unsuprising, as the chiral condensate was used in our calculation, and
$m_{\pi t}$ varies little for $\kappa<\kappa_c$. 

\subsubsection{Pion correlators in the x and y directions}
\label{sec:kren}
\begin{figure}[htb]
\bigskip\bigskip
\begin{center}
\epsfig{file=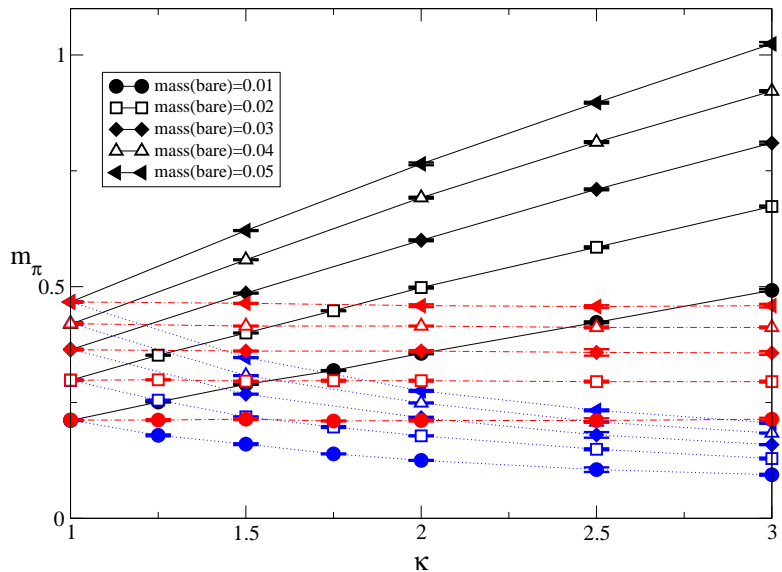, angle=270, width=12cm}
\end{center}
\caption{(Color online) Screening masses $m_{\pi x}$ (solid), $m_{\pi y}$ (dotted), 
and the geometric mean
$\sqrt{m_{\pi x}m_{\pi y}}$ (dot-dashed), versus  $\kappa$.} 
\label{fig:xFit}
\end{figure}
Figure~ \ref{fig:xFit} summarises the variation in the effective pion mass 
in the $x$ and $y$ directions, as obtained by fitting $C_{\pi x,y}$ data to the 
form (\ref{eqn:cosh}).  We were unable to 
fit for $m_{\pi y}$ beyond $\kappa=3.00$; 
the propagator took on a saw-tooth form 
consistent with the pion correlation length being infinite for all practical 
purposes, ie $m_{\pi y}\ll L_y^{-1}$.  Tables \ref{tab:xpimass} and 
\ref{tab:ypimass} give fit windows and $\frac{\chi^{2}}{d.o.f}$ for each point.

The data shows $m_{\pi x}$ increasing with $\kappa$, 
and $m_{\pi y}$ decreasing.  This is not unexpected; naively restoring
explicit factors of lattice spacing we expect $m_{\pi\mu}=M_\pi
a_\mu=\lambda_\mu^{-1}M_\pi a$ where $M_\pi$ is the expected dimensionful pion
mass assuming no physical effect as a result of aniostropy. This implies that 
the mass pole of the propagator is shifted by a factor of 
$\kappa^{\frac{1}{2}}$ in the $x$ direction 
and by $\kappa^{-\frac{1}{2}}$ in the $y$ direction. Indeed, the geometric 
mean $\sqrt{m_{\pi x}m_{\pi y}}$ is approximately
independent of $\kappa$, suggesting
that the results can be explained entirely in terms of equal and opposite
anisotropies, ie. $a_xa_y=a^2$.

\begin{figure}[htb]
\bigskip\bigskip
\begin{center}
\epsfig{file=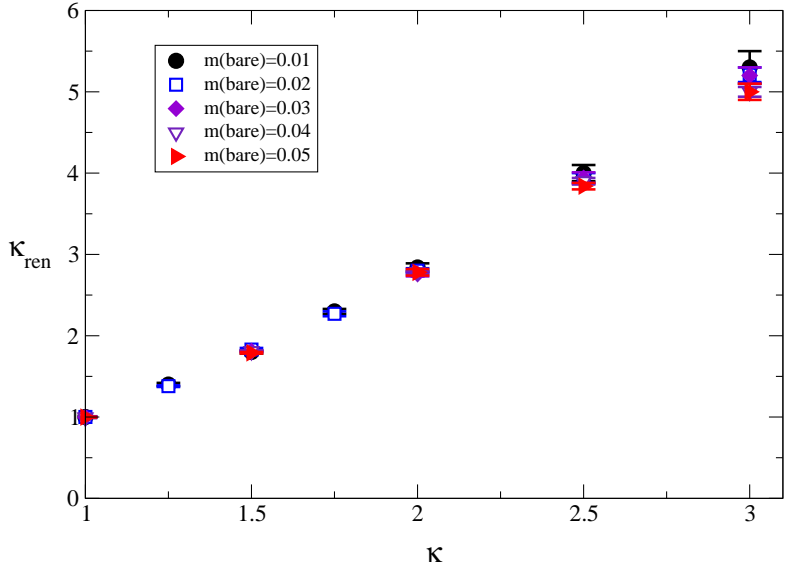,angle=270,width=12cm}
\end{center}
\caption{(Color online) $\kappa_{ren}$ versus $\kappa$ in the accessible regime $\kappa\leq3$.} 
\label{fig:kappar}
\end{figure}
However, we should 
consider the possiblity that as a result of dynamical effects the physical
anisotropy is not simply related to the ``bare'' anisotropy introduced in 
(\ref{eqn:lattact}). We can then regard the ratio $m_{\pi x}/m_{\pi y}$ as a 
measure of the physical or renormalised anisotropy $\kappa_{ren}$
\cite{pionfoot}.  Figure~\ref{fig:kappar} plots 
the resulting 
$\kappa_{ren}(\kappa)$; in fact 
$\kappa_{ren}$ is approximately described  by 
\begin{equation}
{{\kappa_{ren}-1}\over{\kappa-1}}\approx2,
\label{eq:kren}
\end{equation}
a relation which appears remarkably insensitive to the fermion mass $m$.

Eqn.~(\ref{eq:kren}) implies that the physical anisotropy is greater than that
in the bare theory, in direct contradiction to the analytic prediction of 
Lee and Herbut 
\cite{Lee:2002qz}. 
Our results suggest $\kappa_{ren}-1$ is {\em relevant}.
Possible explanations for the discrepancy are 
firstly  that we are not necessarily simulating at a 
small enough anisotropy, or sufficiently close to the continuum limit, where
the results of \cite{Lee:2002qz} apply,  
and secondly, as we have stressed in 
Section~\ref{sec:model}, the model (\ref{eqn:lattact}) does not quite reproduce 
the theory examined there. 

\begin{table}[h]
\centering
\setlength{\tabcolsep}{.85pc}
\renewcommand{\arraystretch}{.9}
\begin{tabular}{|llll|llll|}
\hline
$m$ & $\kappa$ &$\frac{\chi^{2}}{d.o.f}$ & {\small fit window}& 
$m$ & $\kappa$ &$\frac{\chi^{2}}{d.o.f}$ & {\small fit window} \\
\hline
\hline
0.01&	1.00&	1.395&	1-15&	0.02&	1.00&	1.156&	1-15	\\
&	1.25&	0.633&	1-15&	&	1.25&	1.095&	3-13	\\
&	1.50&	1.118&	1-15&	&	1.50&	0.900&	1-15	\\
&	1.75&	0.753&	1-15&	&	1.75&	0.652&	5-11	\\
&	2.00&	1.041&	3-13&	&	2.00&	1.021&	2-14	\\
&	2.50&	1.011&	1-15&	&	2.50&	1.341&	1-15	\\
&	3.00&	1.332&	1-15&	&	3.00&	1.167&	3-13	\\
&	3.50&	0.901&	1-15&	&	3.50&	1.373&	2-14	\\
&	4.00&	1.318&	1-15&	&	4.00&	1.170&	4-12	\\
&	4.50&	1.587&	1-15&	&	4.50&	1.783&	1-15	\\
&	5.00&	1.197&	1-15&	&	5.00&	3.112&	3-13	\\
&	5.50&	1.341&	2-14&	&	5.50&	1.126&	1-15	\\
&	6.00&	1.045&	2-14&	&	6.00&	0.476&	2-14	\\
&	6.50&	1.502&	1-15&	&	6.50&	1.044&	2-14	\\
&	7.00&	1.592&	2-14&	&	7.00&	1.090&	2-14	\\
&	10.00&	1.634&	5-11&	&	10.00&	1.005&	1-15	\\
\hline
0.03&	1.00&	1.080&	2-14&	0.04&	1.00&	1.008&	5-11	\\
&	1.50&	1.014&	4-12&	&	1.50&	0.855&	2-14	\\
&	2.00&	0.961&	1-15&	&	2.00&	1.064&	2-14	\\
&	2.50&	1.298&	3-13&	&	2.50&	0.811&	4-12	\\
&	3.00&	1.065&	1-15&	&	3.00&	1.271&	1-15	\\
&	3.50&	1.058&	3-13&	&	3.50&	0.882&	1-15	\\
&	4.00&	1.112&	5-11&	&	4.00&	1.562&	1-15	\\
&	4.50&	1.684&	2-14&	&	4.50&	1.045&	1-15	\\
&	5.00&	1.202&	1-15&	&	5.00&	0.999&	1-15	\\
&	5.50&	1.301&	1-15&	&	5.50&	0.858&	3-13	\\
&	6.00&	0.804&	1-15&	&	6.00&	1.413&	1-15	\\
&	6.50&	1.166&	1-15&	&	6.50&	1.076&	3-13	\\
&	7.00&	1.071&	1-15&	&	7.00&	1.101&	1-15	\\
&	10.00&	1.128&	4-12&	&	10.00&	0.599&	1-15	\\
\hline
0.05&	1.00&	0.963&	3-13	&&&&\\
&	1.50&	1.368&	4-12	&&&&\\
&	2.00&	1.521&	1-15	&&&&\\
&	2.50&	1.385&	4-12	&&&&\\
&	3.00&	0.914&	5-11	&&&&\\
&	3.50&	0.938&	4-12	&&&&\\
&	4.00&	0.716&	3-13	&&&&\\
&	4.50&	1.544&	4-12	&&&&\\
&	5.00&	1.218&	2-14	&&&&\\
&	5.50&	1.018&	5-11	&&&&\\
&	6.00&	1.003&	5-11	&&&&\\
&	6.50&	0.951&	1-15	&&&&\\
&	7.00&	0.705&	2-14	&&&&\\
&	10.00&	1.893&	2-14	&&&&\\
\hline
\end{tabular}
\caption{Pion mass $m_{\pi t}$ fitting data}
\smallskip
\label{tab:tpimass}
\end{table}

\begin{table}[h]
\centering
\setlength{\tabcolsep}{1.5pc}
\begin{tabular}{|lllll|}
\hline
$m$ & $\kappa$ & $m_{\pi x}$ & $\frac{\chi^{2}}{d.o.f}$ & fit window  \\
\hline
\hline
0.01&	1.00&	0.211(1)&	1.073&	2-14\\
&	1.25&	0.251(1)&	0.927&	1-15\\
&	1.50&	0.289(1)&	0.636&	1-15\\	
&	1.75&	0.319(1)&	0.742&	1-15\\	
&	2.00&	0.356(2)&	1.214&	1-15\\	
&	2.50&	0.423(2)&	0.975&	1-15\\	
&	3.00&	0.492(3)&	1.735&	3-13\\	
\hline
0.02&	1.00&	0.298(1)&	1.256&	1-15\\	
&	1.25&	0.352(1)&	0.812&	1-15\\	
&	1.50&	0.400(1)&	1.016&	1-15\\	
&	1.75&	0.448(1)&	1.397&	2-14\\	
&	2.00&	0.498(2)&	1.000&	2-14\\	
&	2.50&	0.585(2)&	1.313&	2-14\\	
&	3.00&	0.673(2)&	0.709&	1-15\\	
\hline
0.03&	1.00&	0.364(1)&	1.124&	1-15\\	
&	1.50&	0.486(1)&	0.671&	2-14\\	
&	2.00&	0.600(2)&	0.914&	4-12\\
&	2.50&	0.710(2)&	1.115&	1-15\\	
&	3.00&	0.810(3)&	1.074&	5-11\\	
\hline
0.04&	1.00&	0.467(1)&	1.049&	1-15\\	
&	1.50&	0.621(1)&	0.960&	4-12\\	
&	2.00&	0.765(3)&	0.940&	6-10\\	
&	2.50&	0.897(2)&	1.044&	1-15\\	
&	3.00&	1.024(4)&	0.805&	6-10\\	
\hline
0.05&	1.00&	0.419(1)&	1.583&	3-13\\	
&	1.50&	0.558(1)&	0.955&	3-13\\	
&	2.00&	0.692(2)&	0.892&	3-13\\	
&	2.50&	0.812(2)&	0.848&	3-13\\	
&	3.00&	0.922(2)&	0.967&	1-15\\	
\hline
\end{tabular}
\caption{Effective pion mass $m_{\pi x}$ in the $x$ direction}
\smallskip
\label{tab:xpimass}
\end{table}

\begin{table}[h]
\centering
\setlength{\tabcolsep}{1.5pc}
\begin{tabular}{|lllll|}
\hline
$m$ & $\kappa$ & $m_{\pi y}$ & $\frac{\chi^{2}}{d.o.f}$ & fit window  \\
\hline
\hline

0.01&	1.00&	0.212(1)&	1.258&	1-15\\	
&	1.25&	0.179(2)&	1.027&	3-13\\	
&	1.50&	0.160(2)&	0.554&	3-13\\	
&	1.75&	0.139(1)&	0.980&	1-15\\	
&	2.00&	0.125(1)&	1.027&	2-14\\	
&	2.50&	0.105(5)&	1.150&	5-11\\	
&	3.00&	0.094(2)&	0.487&	2-14\\	
\hline
0.02&	1.00&	0.298(1)&	0.598&	1-15\\
&	1.25&	0.255(1)&	1.419&	3-13\\	
&	1.50&	0.219(1)&	0.909&	2-14\\	
&	1.75&	0.197(2)&	0.677&	4-12\\	
&	2.00&	0.178(1)&	1.850&	2-14\\	
&	2.50&	0.149(2)&	0.957&	3-13\\	
&	3.00&	0.129(2)&	1.073&	2-14\\	
\hline
0.03&	1.00&	0.365(1)&	1.648&	1-15\\
&	1.50&	0.268(1)&	0.982&	1-15\\	
&	2.00&	0.217(2)&	0.895&	4-12\\	
&	2.50&	0.180(6)&	1.832&	6-10\\	
&	3.00&	0.159(1)&	1.146&	4-12\\
\hline
0.04&	1.00&	0.421(2)&	1.297&	3-13\\	
&	1.50&	0.308(1)&	1.038&	3-13\\	
&	2.00&	0.249(1)&	1.092&	5-11\\	
&	2.50&	0.208(1)&	2.117&	4-12\\	
&	3.00&	0.184(1)&	0.996&	4-12\\	
\hline
0.05&	1.00&	0.467(1)&	1.034&	3-13\\	
&	1.50&	0.347(1)&	0.921&	3-13\\	
&	2.00&	0.275(2)&	0.779&	5-11\\	
&	2.50&	0.233(2)&	1.069&	4-12\\	
&	3.00&	0.206(2)&	1.089&	4-12\\	
\hline
\end{tabular}
\caption{Effective pion mass $m_{\pi y}$ in the $y$ direction}
\smallskip
\label{tab:ypimass}
\end{table}

\section{Discussion}
\label{sec:discuss}

In this paper we have for the first time presented simulation results for
a condensed matter-inpsired version of 
lattice non-compact QED$_3$, with a physical number of fermion flavours $N_f=2$,
in which anisotropic fermion hopping in the spatial
direction has been explicitly introduced. Our main result is that the
renormalised anisotropy, which we define as the ratio of the pion correlation 
length
in the $y$ direction to that in the $x$, is greater than the bare anisotropy
parameter $\kappa$, and hence that $\kappa$ is a relevant parameter in the 
renormalisation group sense as momentum scales flow towards the infra-red.
Since the ratio $v_F/v_\Delta$ is known to depart from 1 for real compounds, 
this
result implies that apparently universal 
results for, eg. $N_{fc}$ obtained from QED$_3$ in the isotropic limit
$\kappa=1$ must be treated with caution when applied to superconducting
cuprates.

Two caveats should be issued: first, as repeatedly stressed, our model
(\ref{eqn:lattact}) has an overall physical $x - y$ anisotropy, whereas
anisotropies in the
nodal fermion action (\ref{eq:rel}) cancel between flavours 1 and 2.
We have gone some way towards quantifying this effect with our results in 
Fig.~\ref{fig:plq}, which show that the extra anisotropy effects introduced by
our formulation, as manifested by the difference between
$\langle\Theta_{xt}^2\rangle$
and $\langle\Theta_{yt}^2\rangle$, are significantly smaller than the splitting
between these observables and $\langle\Theta_{xy}^2\rangle$, which must 
persist in both models. As discussed in Sec.~\ref{sec:model}, it is possible to
formulate a lattice model with symmetrised aniostropies to check this issue
further, but at the cost of using an inexact simulation algorithm.
Secondly, since as a Goldstone boson 
the pion is a distinguished particle, it is possible that definitions of
$\kappa_{ren}$ in terms correlation lengths of other states may yield a 
different
answer. This will be explored in future work.

An interesting and to some extent unexpected result of our study,
encapsulated in Figs.~\ref{fig:chiralgraph} and \ref{fig:longsusc}, 
is that there
appears to be a chiral symmetry restoring phase transition at
$\kappa_c\simeq4.5$. Strictly speaking, a finite volume scaling study on a range
of lattice volumes will be needed to elucidate the nature of the phase
transition, but if it proves to be second order then it is difficult to avoid
the conclusion that
$\langle\bar\Psi\Psi\rangle=0$ for large anisotropies even for $N_f=2$,
implying that $N_{fc}$ is a decreasing function of $\kappa$,
and that therefore the pseudogap phase persists down to $T=0$ in cuprates.
Physically, the phase fluctuations hypothetically responsible for the 
destruction of superconducting order must then arise as a result of quantum, as
opposed to thermal, effects.

We can estimate the range of values of $\kappa$ for which our results might in
fact be physically relevant.  The empirical equation for the 
boundary of the dSC region \cite{Tallon}
\begin{equation}
\frac{T_c}{T^{\mbox{\scriptsize {\em max}}}_c}=1 - 82.6(p - 0.16)^2,
\end{equation}
used for the cuprate YBCO, where $p$ is the hole concentration, gives
us a value of $p \approx 0.05$ for $T_c=0$ in the underdoped region.
Sutherland {\em et al.} have measured $\kappa$ for four values of the doping of
this cuprate;  from Figure 4 of \cite{Sutherland} one can extrapolate by eye
that $6\lapprox\kappa\lapprox8$ at the onset of the superconducting phase.  
This does seem to
suggest that the QED model 
predicts the occurence of a phase transition at a value
of $\kappa$ within the region of its validity; were $\kappa_c>6$ the phase
transition would occur after the onset of superconductivity, which would be
unphysical.

Finally, it is interesting to speculate on the nature of the chirally symmetric
high-$\kappa$ phase. Franz {\em et al\/} \cite{Franz:2002qy} have argued that 
in the chirally symmetric phase the fermion propagator $\langle
\Psi(0)\bar\Psi(x)\rangle$ receives a large anomalous scaling dimension from
quantum corrections, which are calculable as a power series in $N_f^{-1}$.
We find it difficult to reconcile these ideas with the plaquette data
of Fig.~\ref{fig:plq} which show plaquette fluctuations increasing 
with $\kappa$,
whereas massless fermions would be expected to suppress such fluctuations
through screening. A plausible alternative is that chiral symmetry restoration
in the model is itself driven by fluctuations of the phase of 
$\bar\Psi\Psi$ as the system dynamics becomes more and more two dimensional
with increasing $\kappa$, but that fermion mass generation, which depends
on $\langle\vert\bar\Psi\Psi\vert\rangle$, remains insensitive to $\kappa$.
Similar effects have been observed in model 
simulations in which phase fluctuations 
are thermally driven \cite{Hands:2001cs}. In future work we intend to explore
this issue further with measurements of the gauge-fixed fermion propagator.

\section*{Acknowledgements}
SJH is supported by a PPARC Senior Research Fellowship, and IOT by a University
of Wales Research Student Scholarship. We have greatly benefitted from
discussions with Chris Allton, Igor Herbut and Zlatko Te\v sanovi\'c.


\begin{thebibliography}{xx}
\bibitem{chiralfoot}Strictly speaking, in the continuum limit
chiral symmetry is enlarged to U($2N_f$) which spontaneously breaks to
U($N_f)\otimes$U($N_f$) implying a total of $2N_f^2$ Goldstones, some of which
are scalar \cite{Hands:2004bh}; since we are
far from the continuum limit we ignore this complication.
\bibitem{pionfoot}Strictly speaking, this only refers to the effect 
of the anisotropy on the pion, a Goldstone boson of the system.  
It remains to be seen if this is true of other more generic states.
\end{thebibliography}
\end{document}